\documentclass[12pt,a4paper]{iopart}
\renewcommand\etal{\textit{et al}}

\expandafter\let\csname equation*\endcsname\relax
\expandafter\let\csname endequation*\endcsname\relax

\usepackage[UKenglish]{babel}
\usepackage[utf8]{inputenc}
\usepackage{graphicx}
\usepackage{xparse}
\usepackage{amsmath}
\usepackage{amssymb}
\usepackage{mathtools}
\usepackage{subdepth}
\usepackage{dsfont}
\usepackage{hyperref}
\usepackage{glossaries}
\usepackage[noabbrev]{cleveref}
\usepackage{csquotes}
\usepackage{cite}

\crefname{equation}{}{}
\crefname{table}{table}{tables}
\crefname{figure}{figure}{figures}
\crefname{section}{section}{sections}

\Crefname{equation}{Equation}{Equations}
\Crefname{table}{Table}{Tables}
\Crefname{figure}{Figure}{Figures}
\Crefname{section}{Section}{Sections}

\crefalias{subequation}{equation}
\crefalias{subsection}{section}

\newcommand*\subrefformat[2]{#1~(#2)}

\NewDocumentCommand\subref{smm}{%
  \subrefformat{\IfBooleanTF{#1}{\ref*{#2}}{\ref{#2}}}{#3}%
}

\NewDocumentCommand\csubref{smm}{%
  \subrefformat{\IfBooleanTF{#1}{\cref*{#2}}{\cref{#2}}}{#3}%
}

\NewDocumentCommand\csubrefs{smm}{%
  \subrefformat{\IfBooleanTF{#1}{\namecrefs*{#2} \labelcref{#2}}%
    {\namecrefs{#2} \labelcref{#2}}}{#3}%
}

\NewDocumentCommand\Csubref{smm}{%
  \subrefformat{\IfBooleanTF{#1}{\Cref*{#2}}{\Cref{#2}}}{#3}%
}

\NewDocumentCommand\Csubrefs{smm}{%
  \subrefformat{\IfBooleanTF{#1}{\nameCrefs*{#2} \labelcref{#2}}%
    {\nameCrefs{#2} \labelcref{#2}}}{#3}%
}

\NewDocumentCommand\labelcsubref{smm}{%
  \subrefformat{\IfBooleanTF{#1}{\labelcref*{#2}}{\labelcref{#2}}}{#3}%
}

\let\stdphi\phi%
\let\phi\varphi%
\let\varphi\stdphi%

\DeclareMathAlphabet\mathbfcal{OMS}{cmsy}{b}{n}

\NewDocumentCommand\id{}{\mathds{1}}

\NewDocumentCommand\realnumbers{}{\mathds{R}}
\NewDocumentCommand\complexnumbers{}{\mathds{C}}

\NewDocumentCommand\comma{}{\,,}
\NewDocumentCommand\period{}{\,.}
\NewDocumentCommand\imag{}{\mathrm{i}\mkern1mu} 

\NewDocumentCommand\conjg{sgod()}{
  \IfValueTF{#2}
  {#2^\ast}
  {
    \IfValueTF{#3}
    {\IfBooleanTF{#1}{[#3]^\ast}{\qty[#3]^\ast}}
    {
      \IfValueT{#4}
      {\IfBooleanTF{#1}{(#4)^\ast}{\qty(#4)^\ast}}
    }
  }
}
\NewDocumentCommand\adjg{sgod()}{
  \IfValueTF{#2}
  {#2^\dagger}
  {
    \IfValueTF{#3}
    {\IfBooleanTF{#1}{[#3]^\dagger}{\qty[#3]^\dagger}}
    {
      \IfValueT{#4}
      {\IfBooleanTF{#1}{(#4)^\dagger}{\qty(#4)^\dagger}}
    }
  }
}
\NewDocumentCommand\inv{sgod()}{
  \IfValueTF{#2}
  {#2^{-1}}
  {
    \IfValueTF{#3}
    {\IfBooleanTF{#1}{[#3]^{-1}}{\qty[#3]^{-1}}}
    {
      \IfValueT{#4}
      {\IfBooleanTF{#1}{(#4)^{-1}}{\qty(#4)^{-1}}}
    }
  }
}
\NewDocumentCommand\transp{sgod()}{
  \IfValueTF{#2}
  {#2^\intercal}
  {
    \IfValueTF{#3}
    {\IfBooleanTF{#1}{[#3]^\intercal}{\qty[#3]^\intercal}}
    {
      \IfValueT{#4}
      {\IfBooleanTF{#1}{(#4)^\intercal}{\qty(#4)^\intercal}}
    }
  }
}

\NewDocumentCommand\rmR{o}{\IfNoValueTF{#1}{\mathrm{R}}{\mathrm{R}_{#1}}}
\NewDocumentCommand\rmL{o}{\IfNoValueTF{#1}{\mathrm{L}}{\mathrm{L}_{#1}}}

\NewDocumentCommand\op{sm}
{\IfBooleanTF{#1}
  {\ensuremath{\hat{#2}}}
  {\ensuremath{\mathcal{#2}}}
}

\NewDocumentCommand\lsym{o}{\ensuremath{\op{S}_\mathrm{L\IfValueT{#1}{, #1}}}}
\NewDocumentCommand\rsym{o}{\ensuremath{\op{S}_\mathrm{R\IfValueT{#1}{, #1}}}}

\NewDocumentCommand\tpsi{}{\tilde{\psi}}

\NewDocumentCommand\ac{m}{\gls*{#1}}

\NewDocumentCommand\acrodef{omm}{
  \IfNoValueTF{#1}{
    \newacronym{#2}{#2}{#3}
  }{
    \newacronym[#1]{#2}{#2}{#3}
  }
}

\acrodef{QM}{quantum mechanics}

\NewDocumentCommand\abs{ l m }{\braces#1{\lvert}{\rvert}{#2}} 
\NewDocumentCommand\norm{ l m }{\braces#1{\lVert}{\rVert}{#2}} 
\NewDocumentCommand\flatfrac{ m m }{\left.#1\middle\slash#2\right.}
\RenewDocumentCommand\Re{g}{\IfNoValueTF{#1}{\operatorname{Re}}{\fbraces{\lbrace}{\rbrace}{\operatorname{Re}}{#1}}}
\RenewDocumentCommand\Im{g}{\IfNoValueTF{#1}{\operatorname{Im}}{\fbraces{\lbrace}{\rbrace}{\operatorname{Im}}{#1}}}

\NewDocumentCommand\opbraces{ m g o d() }
{
  \IfNoValueTF{#2}
  {
    \IfNoValueTF{#3}
      {
        \IfNoValueTF{#4}
        {#1}
        {\fbraces{\lparen}{\rparen}{#1}{#4}}
      }
      {
        \fbraces{\lbrack}{\rbrack}{#1}{#3}
        \IfNoValueTF{#4}{}{(#4)}
      }
  }
  {
    \fbraces{\lbrace}{\rbrace}{#1}{#2}
    \IfNoValueTF{#3}{}{[#3]}
    \IfNoValueTF{#4}{}{(#4)}
  }
}
\NewDocumentCommand\trigbraces{ m o d() }
{
  \IfNoValueTF{#3}
  {#1 \IfNoValueTF{#2}{}{[#2]}}
  {#1 \IfNoValueTF{#2}{}{^{#2}} \argopen(#3\argclose)}
}

\DeclareMathOperator{\rank}{rank} 

\let\exponential\exp \RenewDocumentCommand\exp{}{\opbraces{\exponential}}
\let\logarithm\log \RenewDocumentCommand\log{}{\trigbraces{\logarithm}}
\let\naturallogarithm\ln \RenewDocumentCommand\ln{}{\trigbraces{\naturallogarithm}}
\let\determinant\det \RenewDocumentCommand\det{}{\opbraces{\determinant}}

\makeatletter
\NewDocumentCommand\qty{}{{\ifnum\z@=`}\fi\@quantity}
\NewDocumentCommand\@quantity{ t\big t\Big t\bigg t\Bigg g o d() d|| }
{ 
  \IfBooleanTF{#1}{\let\ltag\bigl \let\rtag\bigr}{
    \IfBooleanTF{#2}{\let\ltag\Bigl \let\rtag\Bigr}{
      \IfBooleanTF{#3}{\let\ltag\biggl \let\rtag\biggr}{
        \IfBooleanTF{#4}
        {\let\ltag\Biggl \let\rtag\Biggr}
        {\let\ltag\left \let\rtag\right}
      }
    }
  }
  \IfNoValueTF{#5}{
    \IfNoValueTF{#6}{
      \IfNoValueTF{#7}{
        \IfNoValueTF{#8}
        {()}
        {\ltag\lvert{#8}\rtag\rvert}
      }
      {\ltag(#7\rtag) \IfNoValueTF{#8}{}{|#8|}}
    }
    {\ltag[#6\rtag] \IfNoValueTF{#7}{}{(#7)} \IfNoValueTF{#8}{}{|#8|}}
  }
  {\ltag\lbrace#5\rtag\rbrace  \IfNoValueTF{#6}{}{[#6]} \IfNoValueTF{#7}{}{(#7)} \IfNoValueTF{#8}{}{|#8|}}
  \ifnum\z@=`{\fi}
}
\makeatother

\NewDocumentCommand\pmqty{m}{\begin{pmatrix}#1\end{pmatrix}}
\NewDocumentCommand\Pmqty{m}{\left\lgroup\begin{matrix}#1\end{matrix}\right\rgroup}
\NewDocumentCommand\bmqty{m}{\begin{bmatrix}#1\end{bmatrix}}
\NewDocumentCommand\vmqty{m}{\begin{vmatrix}#1\end{vmatrix}}

\makeatletter
\NewDocumentCommand\mqty{}{{\ifnum\z@=`}\fi\@matrixquantity}
\NewDocumentCommand\@matrixquantity{ s g o d() d|| }
{
  \mathord{
  \IfNoValueTF{#2}
  {
    \IfNoValueTF{#3}
    {
      \IfNoValueTF{#4}
      {
        \IfNoValueTF{#5}
        {()}
        {\vmqty{#5}}
      }
      {
        \IfBooleanTF{#1}
        {\Pmqty{#4}}
        {\pmqty{#4}}
        \IfNoValueTF{#5}{}{|#5|}
      }
    }
    {\bmqty{#3} \IfNoValueTF{#4}{}{(#4)} \IfNoValueTF{#5}{}{|#5|}}
  }
  {\begin{matrix}#2\end{matrix} \IfNoValueTF{#3}{}{[#3]} \IfNoValueTF{#4}{}{(#4)} \IfNoValueTF{#5}{}{|#5|}}
  }
  \ifnum\z@=`{\fi}
}
\makeatother

\NewDocumentCommand\dv{ s o m g d() }
{ 
  \IfBooleanTF{#1}
  {\let\fractype\flatfrac}
  {\let\fractype\frac}
  \IfNoValueTF{#4}
  {
    \IfNoValueTF{#5}
    {\fractype{\diffd \IfNoValueTF{#2}{}{^{#2}}}{\diffd #3\IfNoValueTF{#2}{}{^{#2}}}}
    {\fractype{\diffd \IfNoValueTF{#2}{}{^{#2}}}{\diffd #3\IfNoValueTF{#2}{}{^{#2}}} \argopen(#5\argclose)}
  }
  {\fractype{\diffd \IfNoValueTF{#2}{}{^{#2}} #3}{\diffd #4\IfNoValueTF{#2}{}{^{#2}}}}
}

\NewDocumentCommand\pdv{ s o m g g d() }
{ 
  \IfBooleanTF{#1}
  {\let\fractype\flatfrac}
  {\let\fractype\frac}
  \IfNoValueTF{#4}
  {
    \IfNoValueTF{#6}
    {\fractype{\partial \IfNoValueTF{#2}{}{^{#2}}}{\partial #3\IfNoValueTF{#2}{}{^{#2}}}}
    {\fractype{\partial \IfNoValueTF{#2}{}{^{#2}}}{\partial #3\IfNoValueTF{#2}{}{^{#2}}} \argopen(#6\argclose)}
  }
  {
    \IfNoValueTF{#5}
    {\fractype{\partial \IfNoValueTF{#2}{}{^{#2}} #3}{\partial #4\IfNoValueTF{#2}{}{^{#2}}}}
    {\fractype{\partial^2 #3}{\partial #4 \partial #5}}
  }
}

\makeatletter
\NewDocumentCommand\braces{}{{\ifnum\z@=`}\fi\@braces}
\NewDocumentCommand\@braces{ s t\big t\Big t\bigg t\Bigg m m m }
{ 
  \IfBooleanTF{#1}
  {\left#6\smash{#8}\right#7\vphantom{#8}}
  {
    \IfBooleanTF{#2}{\bigl#6{#8}\bigr#7}{
      \IfBooleanTF{#3}{\Bigl#6{#8}\Bigr#7}{
        \IfBooleanTF{#4}{\biggl#6{#8}\biggr#7}{
          \IfBooleanTF{#5}{\Biggl#6{#8}\Biggr#7}{\left#6{#8}\right#7}
        }
      }
    }
  }
  \ifnum\z@=`{\fi}
}
\makeatother
\NewDocumentCommand\commutator{ l m m }{\braces#1{\lbrack}{\rbrack}{#2,#3}} 
\NewDocumentCommand\anticommutator{ l m m }{\braces#1{\lbrace}{\rbrace}{#2,#3}} 

\NewDocumentCommand\bra{ s m t\ket s g }
{ 
  \IfBooleanTF{#3}
  { 
    \IfBooleanTF{#1}
    { 
      \IfNoValueTF{#5}
      {\braket*{#2}{} \IfBooleanTF{#4}{*}{}}
      {\braket*{#2}{#5}}
    }
    {
      \IfBooleanTF{#4}
      { 
        \IfNoValueTF{#5}
        {\braket{#2}{} *}
        {\braket*{#2}{#5}}
      }
      {\braket{#2}{\IfNoValueTF{#5}{}{#5}}} 
    }
  }
  { 
    \IfBooleanTF{#1}
    {\vphantom{#2}\left\langle\smash{#2}\right\rvert}
    {\left\langle{#2}\right\rvert}
    \IfBooleanTF{#4}{*}{}
    \IfNoValueTF{#5}{}{#5}
  }
}

\NewDocumentCommand\ket{ s m }
{ 
  \IfBooleanTF{#1}
  {\vphantom{#2}\left\lvert\smash{#2}\right\rangle} 
  {\left\lvert{#2}\right\rangle} 
}

\NewDocumentCommand\braket{ s m g }
{ 
  \IfBooleanTF{#1}
  { 
    \IfNoValueTF{#3}
    {\vphantom{#2}\left\langle\smash{#2}\middle\vert\smash{#2}\right\rangle}
    {\vphantom{#2#3}\left\langle\smash{#2}\middle\vert\smash{#3}\right\rangle}
  }
  { 
    \IfNoValueTF{#3}
    {\left\langle{#2}\middle\vert{#2}\right\rangle}
    {\left\langle{#2}\middle\vert{#3}\right\rangle}
  }
}

\NewDocumentCommand\ketbra{ s m g }
{ 
  \IfBooleanTF{#1}
  { 
    \IfNoValueTF{#3}
    {\vphantom{#2}\left\lvert\smash{#2}\middle\rangle\!\middle\langle\smash{#2}\right\rvert}
    {\vphantom{#2#3}\left\lvert\smash{#2}\middle\rangle\!\middle\langle\smash{#3}\right\rvert}
  }
  { 
    \IfNoValueTF{#3}
    {\left\lvert{#2}\middle\rangle\!\middle\langle{#2}\right\rvert}
    {\left\lvert{#2}\middle\rangle\!\middle\langle{#3}\right\rvert}
  }
}

\usepackage{tensor}

\NewDocumentCommand\bileft{ m }{\mathopen{}\mathclose\bgroup\left#1}
\NewDocumentCommand\biright{ m }{\aftergroup\egroup\right#1}
\NewDocumentCommand\bitensor{ o m o }
{\left.\kern-\nulldelimiterspace\IfNoValueTF{#1}
  {
    \IfNoValueTF{#3}
    {\tensor*[]{#2}{}}
    {\tensor*[]{#2}{_{\mathrm{#3}}}}
  }{
    \IfNoValueTF{#3}
    {\tensor*[_{\mathrm{#1}}]{#2}{}}
    {\tensor*[_{\mathrm{#1}}]{#2}{_{\mathrm{#3}}}}
  }\right.\kern-\nulldelimiterspace}

\NewDocumentCommand\biket{ m s m }
{ 
  \bitensor{
    \IfBooleanTF{#2}
    {\vphantom{#3}\bileft\lvert\smash{#3}\biright\rangle} 
    {\bileft\lvert{#3}\biright\rangle} 
  }[#1]
}

\NewDocumentCommand\lket{}{\biket{L}}
\NewDocumentCommand\rket{}{\biket{R}}

\NewDocumentCommand\bibra{ m s m }
{ 
  \bitensor[#1]{
    \IfBooleanTF{#2}
    {\vphantom{#3}\bileft\langle\smash{#3}\biright\rvert} 
    {\bileft\langle{#3}\biright\rvert} 
  }
}

\NewDocumentCommand\lbra{}{\bibra{L}}
\NewDocumentCommand\rbra{}{\bibra{R}}

\NewDocumentCommand\bibraket{ m m s m g g }
{ 
  \bitensor[#1]{
    \IfBooleanTF{#3}
    { 
      \IfNoValueTF{#5}
      {\vphantom{#4}\langle\smash{#4}\vert\smash{#4}\biright\rangle}
      {
        \IfNoValueTF{#6}
        {\vphantom{#4#5}\langle\smash{#4}\vert\smash{#5}\rangle}
        {\vphantom{#4#5#6}\langle\smash{#4}\vert\smash{#5}\vert\smash{#6}\rangle}
      }
    }
    { 
      \IfNoValueTF{#5}
      {\bileft\langle{#4}\middle\vert{#4}\biright\rangle}
      {
        \IfNoValueTF{#6}
        {\bileft\langle{#4}\middle\vert{#5}\biright\rangle}
        {\vphantom{#5}\bileft\langle{#4}\middle\vert\smash{#5}\middle\vert{#6}\biright\rangle}
      }
    }
  }[#2]
}

\NewDocumentCommand\lrbraket{}{\bibraket{L}{R}}

\NewDocumentCommand\biketbra{ m m s m g g }
{ 
  \bitensor[#1]{
    \IfBooleanTF{#3}
    { 
      \IfNoValueTF{#5}
      {\vphantom{#4}\lvert\smash{#4}\rangle\!\langle\smash{#4}\rvert}
      {
        \IfNoValueTF{#6}
        {\vphantom{#4}\lvert\smash{#4}\rangle\!\langle\smash{#5}\rvert}
        {\vphantom{#4#5#6}\lvert\smash{#4}\rangle\smash{#5}\langle\smash{#6}\rvert}
      }
    }
    { 
      \IfNoValueTF{#5}
      {\bileft\lvert{#4}\middle\rangle\!\middle\langle{#4}\biright\rvert}
      {
        \IfNoValueTF{#6}
        {\bileft\lvert{#4}\middle\rangle\!\middle\langle{#5}\biright\rvert}
        {\vphantom{#5}\bileft\lvert{#4}\middle\rangle\smash{#5}\middle\langle{#6}\biright\rvert}
      }
    }
  }[#2]
}

\NewDocumentCommand\llketbra{}{\biketbra{L}{L}}

\NewDocumentCommand\rrketbra{}{\biketbra{R}{R}}

\NewDocumentCommand\lketbra{}{\llketbra}
\NewDocumentCommand\rketbra{}{\rrketbra}

\begin{document}

\title{Balancing gain and loss in symmetrised multi-well potentials}

\author{Daniel Dizdarevic$^1$, Holger Cartarius$^2$, J\"org Main$^1$, G\"unter Wunner$^1$}

\address{$^1$ Institut f\"ur Theoretische Physik 1, Universit\"at Stuttgart, 70550 Stuttgart, Germany}
\address{$^2$ Physik und ihre Didaktik, Universit\"at Stuttgart, 70550 Stuttgart, Germany}
\ead{daniel.dizdarevic@itp1.uni-stuttgart.de}
\vspace{10pt}
\begin{indented}
\item[] \today
\end{indented}


\begin{abstract}
Balanced gain and loss leads to stationary dynamics in open systems. This occurs naturally in $\mathcal{PT}$-symmetric systems, where the imaginary part of the potential describing gain and loss is perfectly antisymmetric. While this case seems intuitive, stationary dynamics are also possible in asymmetric open systems. Open multi-well quantum systems can possess completely or partly real spectra if their Hamiltonian is symmetrised or semi-symmetrised, respectively. In contrast to similar concepts, symmetrisation allows for the description of physical multi-well potentials with gain and loss. A simple matrix model for the description of two and three-mode systems is used as an example, for which analytical symmetrised solutions are derived. It is explicitly shown how symmetrisation can be used to systematically find two-mode systems with a stable, stationary ground state and why only $\mathcal{PT}$-symmetric two-mode systems can have stationary excited states.
\end{abstract}


%
%
%

\section{Introduction}

Non-Hermitian Hamiltonians have been used in physics since the emergence of \ac{QM} to solve certain problems which are either not solvable within the framework of ordinary \ac{QM} or only with great difficulty \cite{Moiseyev2011}. They are, for example, particularly suited to effectively describe open quantum systems (see e.\,g.\ \cite{Graefe2008a}), which may be arbitrarily hard to treat by explicitly taking the environment into account. However, the interest in non-Hermitian Hamiltonians rapidly increased after Bender and Boettcher introduced the concept of $\op{PT}$ symmetry in 1998 \cite{Bender1998}. $\op{PT}$ symmetry means that a quantum system is invariant with respect to combined parity and time reflections. The outstanding feature of $\op{PT}$-symmetric open quantum systems is that their eigenvalue spectra can be entirely real, even though the Hamiltonian is non-Hermitian, or consist of complex conjugate pairs otherwise. However, $\op{PT}$ symmetry requires a symmetric real potential and an antisymmetric imaginary potential, which is difficult to realise in an experiment \cite{Altinisik2019}.

$\op{PT}$ symmetry is well established nowadays and several books were published on the subject \cite{Christodoulides2018,Bender2019}. Since its first observation in optical wave guides \cite{Guo2009}, $\op{PT}$ symmetry was also observed, among other fields, in mechanical \cite{Bender2013,Bender2019}, electrical \cite{Schindler2011,Schindler2012}, and only recently in quantum systems \cite{Li2019a,Wu2019,Naghiloo2019}. In recent years, advances were made towards technical applications in superconducting wires \cite{Rubinstein2007,Chtchelkatchev2012}, $\op{PT}$ lasers \cite{Chong2011,Liertzer2012}, synthetic materials \cite{Regensburger2012,Castaldi2013}, NMR spectroscopy \cite{Zheng2013}, and even in wireless power transfer \cite{Assawaworrarit2017,Sakhdari2020}.

The concept of $\op{PT}$ symmetry is intriguingly powerful, yet simple and versatile in application. It must be noted, however, that already several years before its introduction, Scholtz, Geyer, and Hahne published a paper on \emph{quasi-Hermitian Hamiltonians in non-Hermitian \ac{QM}} \cite{Scholtz1992}. Quasi-Hermiticity does not rely on the strict symmetry conditions on the potentials which $\op{PT}$ symmetry requires, but allows for the occurrence of real eigenvalues in asymmetric potentials and even in cases with either pure gain or loss \cite{Nixon2016a}. The latter was experimentally observed a while ago in anti-$\op{PT}$-symmetric systems \cite{Ge2013,Wu2014}. In a similar fashion as $\op{PT}$ symmetry, the applicability of quasi-Hermitian \ac{QM} ranges from scattering problems \cite{Sternheim1972,Ruschhaupt2018,Simon2019} to constant-intensity waves \cite{Makris2015,Brandstoetter2019}, which were recently experimentally realised with pressure waves \cite{Rivet2018}. Quasi-Hermiticity can also be used to define a generalised entropy functional for non-Hermitian quantum systems \cite{Stenholm2002,Jakob2003}.

There are other types of non-$\op{PT}$-symmetric systems with similar properties. Anti-$\op{PT}$ symmetry was, for example, observed for coupled atomic spin waves \cite{Peng2016}, in electrical circuits \cite{Choi2018}, and in diffusive systems \cite{Li2019b}. Another type of non-$\op{PT}$-symmetric potentials is based on the connection between the Zakharov–Shabat spectral problem and the Schr\"odinger eigenvalue problem \cite{Wadati2008,Tsoy2014,Konotop2014,Nixon2016a,Nixon2016b,Yang2017} which, for example, allows for the construction of unidirectionally invisible asymmetric potentials \cite{Brandstoetter2019}.

The characteristic feature of such non-$\op{PT}$-symmetric potentials is clearly their lack of any obvious symmetry. Hence they can be applied to situations where gain or loss are either uncontrolled or even uncontrollable. This could be useful for quantum transport in chain potentials with gain and loss. Such systems were recently discussed in the context of $\op{PT}$-symmetric quantum dot chains \cite{Zhang2019}. However, the realisation of stable $\op{PT}$-symmetric potentials is demanding and small perturbations will immediately break the symmetry. A first step towards applications using Bose-Einstein condensates with asymmetric potentials has been made by Lunt \etal{} \cite{Lunt2017}, who reported on the formation of a steady ground state in a non-$\op{PT}$-symmetric, two-mode Bose-Einstein condensate with balanced gain and loss. Due to the non-linear properties of the condensate the system is stable with respect to small asymmetries in gain and loss. However, with their approach the occurrence of steady states seems phenomenological. The aim of the present paper is to continue this line of thought and to describe the underlying mechanism to systematically find asymmetric potentials with balanced gain and loss with discrete complex-conjugated spectra. This allows applications to higher-order multi-well potentials and even continuous systems \cite{Altinisik2019}.

The paper is organised as follows. In the first part we give a short and rather general overview of the concept of \emph{symmetrisation}, which leads to a complex-conjugated eigenvalue structure. However, in contrast to other works (see e.\,g.\ \cite{Mostafazadeh2010} and references therein) we consistently use right-handed and left-handed representations, which naturally occur in non-Hermitian \ac{QM} and allow for a generalisation we call semi-symmetrisation. \Cref{sub:symmetrisation} introduces the basic formalism of symmetrisation in non-Hermitian \ac{QM} and establish its connection to the presence of an antiunitary symmetry. We then discuss its relations to similar concepts and the physical point of view used throughout this paper. In \cref{sub:semi_symmetrisation} the generalisation to semi-symmetrisable systems is introduced, which allows for the occurrence of isolated complex resonances in the spectra. This is of vital importance to interpret the results in \cref{sec:applications}. \Cref{sub:physical_complex_potentials} deals with the physical complex potentials allowed by different approaches. This discussion connects the mathematical theory with the physical systems investigated in the second part of this paper. Finally, in \cref{sec:applications} the concept of symmetrisation is applied to linear two-mode and three-mode quantum systems which are described by a simple few-mode matrix model corresponding to the tight-binding approximation of the Schrödinger equation in a complex few-well potential \cite{Trombettoni2001,Smerzi2003,Kreibich2014}. Such matrix models can also be understood as the mean-field limit of many-body systems \cite{Dizdarevic2018} being good approximations even in the presence of complex potentials \cite{Graefe2008a,Graefe2008b,Graefe2010}. It is shown that a two-mode open quantum system can either only be $\op{PT}$-symmetric or semi-symmetrisable. Thus, the seemingly coincidental occurrence of isolated real eigenvalues in \cite{Lunt2017} can be explained with semi-symmetrisation. Systems with three or more modes can be symmetrised completely. This allows for the occurrence of second and third-order exceptional points. Last but not least an anti-$\op{PT}$-symmetric potential serves as an example for a completely symmetrised three-mode system.

\section{Theory}%
\label{sec:theory}

In the following we discuss under which circumstances a non-Hermitian Hamiltonian has a spectrum which consists either entirely or partly of real or complex-conjugated eigenvalues. We start with an intuitive motivation for the concept of symmetrisation and make connections to the fundamental symmetries in \ac{QM} and similar concepts. Later, we discuss physical potentials allowed by this concept.

\subsection{Symmetrisation}%
\label{sub:symmetrisation}

Basically, there exist two possible approaches in the literature to ensure that the spectrum of an operator is complex conjugate to itself, namely an antiunitary symmetry, which is the statement of Wigner's theorem \cite{Wigner1931,Simon2018}, and symmetrisation. Since these approaches are essential for the present paper, we give a short review of this topic. For the sake of simplicity and clarity we assume that the spectra considered are discrete and non-degenerate. The case with degeneracy is, for example, described in \cite{Solombrino2002}.

\paragraph*{Antiunitary symmetry:}

Consider a non-Hermitian operator $\op{H}$ and the corresponding right-hand Schr\"odinger equation
\begin{equation}
  \op{H} \rket{\psi_n} = E_n \rket{\psi_n}
  \comma
  \label{eq:right_eigenvalue_eq}
\end{equation}
where $E_n$ are the complex eigenvalues of $\op{H}$ with right eigenstates $\rket{\psi_n}$. By introducing an \emph{antilinear} operator $\op{A}$, where antilinear means
\begin{equation}
  \op{A} \qty(\alpha \ket{\alpha} + \beta \ket{\beta})
    = \conjg{\alpha} \op{A} \ket{\alpha} + \conjg{\beta} \op{A} \ket{\beta}
  \comma
\end{equation}
one finds a right-hand eigenvalue equation $\op{H} \rket*{\tpsi_n} = \conjg{E}_n \rket*{\tpsi_n}$ with $\rket*{\tpsi_n} = \op{A} \rket{\psi_n}$ if
\begin{equation}
  \commutator{\op{A}}{\op{H}} = 0
  \period
  \label{eq:antiunitary_symmetry}
\end{equation}
Therefore, if $E_n \in \complexnumbers$ is in the spectrum of $\op{H}$ so is $\conjg{E}_n$; the eigenvalues arise in complex-conjugate pairs. Note that this is in particular true for real eigenvalues \cite{Wigner1960}, which can be considered as their own complex conjugates.

In physical terms this antiunitary symmetry can be related to time-reversal invariance \cite{Sternheim1972}, as the time-reversal operator is antilinear \cite{Wigner1931}. This does, however, not exclude more general antiunitary symmetries, of which $\op{PT}$ symmetry \cite{Bender1998} is a prominent recent example. It is well known that the spectra of $\op{PT}$-symmetric Hamiltonians, i.\,e.\ which satisfy $\commutator{\op{H}}{\op{PT}} = 0$, are in general complex conjugate and can thus also be partially or entirely real \cite{Bender1998,Bender2007}.

\paragraph*{Symmetrisation:}

Instead of \cref{eq:right_eigenvalue_eq} we now consider its adjoint equation. The complex conjugate of $E_n$ then appears naturally on the right-hand side and we can introduce a \emph{linear} operator $\lsym$ in such a way that
\begin{equation}
  \rbra{\psi_n} \adjg{\op{H}} \adjg{\lsym} = \rbra{\psi_n} \adjg{\lsym} \conjg{E}_n
  \comma
  \label{eq:left_eigenvalue_eq_lsym}
\end{equation}
where $\dagger$ indicates the usual Hermitian adjoint defined by $\braket*{\op{H} \varphi}{\psi} = \braket*{\varphi}{\adjg{\op{H}} \psi}$ using the standard Hermitian form. Requiring that the operator $\lsym$ satisfies the relation
\begin{equation}
  \lsym \op{H} = \adjg{\op{H}} \lsym
  \comma
  \label{eq:left_symmetrisation}
\end{equation}
we find a \emph{left-hand} eigenvalue equation
\begin{equation}
  \lbra*{\tpsi_n} \op{H} = \lbra*{\tpsi_n} \conjg{E}_n
  \label{eq:left_eigenvalue_eq}
\end{equation}
with a left eigenstate $\lket*{\tpsi_n} = \lsym \rket*{\psi_n} \neq 0$, which must not necessarily be normalised. Since the Hamiltonian considered is not Hermitian, its eigenbasis is bi-orthogonal (cf.\ \cite{Sternheim1972,Moiseyev2011,Brody2013}). This means that the left and right eigenstates of the same eigenvalue do not necessarily coincide, as it would be the case for Hermitian operators.

\Cref{eq:left_symmetrisation} shows that the combination of $\op{H}$ and $\lsym$ is Hermitian if $\lsym$ is Hermitian. A Hamiltonian satisfying \cref{eq:left_symmetrisation} is thus called \emph{symmetrised} (see \cite{Reid1951,Stenholm2002}) with respect to the left symmetrisation operator $\lsym$. This corresponds to the method proposed by Darboux \cite{Darboux1882,Cannata1998}, which relates the spectra of the Hamiltonians $\op{H}$ and $\adjg{\op{H}}$. The left symmetrisation operator can also be understood as a symmetry of $\op{H}$ within a superoperator framework \cite{Simon2018}. It transforms a right eigenstate of $\op{H}$ into a left eigenstate of $\op{H}$ with the complex conjugate eigenvalue.

By introducing the antilinear operator
\begin{equation}
  \op{T}_\mathrm{L} = \sum_m \lketbra{\psi_m}{\op{K}}{\psi_m}
  \comma
  \label{eq:left_antiunitary_op}
\end{equation}
where $\op{K}$ is the operator associated with complex conjugation, one finds that
\begin{equation}
  \lsym = \op{T}_\mathrm{L} \op{A}
  \label{eq:unitary_left_symmetrisation_op}
\end{equation}
is linear and solves \cref{eq:left_symmetrisation} if $\op{A}$ satisfies \cref{eq:antiunitary_symmetry}. It is a simple exercise to check that $\op{T}_\mathrm{L}$ also satisfies the relation \cref{eq:left_symmetrisation}, that is $\op{T}_\mathrm{L} \op{H} = \adjg{\op{H}} \op{T}_\mathrm{L}$. This shows that symmetrisation of $\op{H}$ is equivalent to generalised $\op{PT}$ symmetry with some linear operator $\op{P} \equiv \lsym$ and some antilinear operator $\op{T} \equiv \op{T}_\mathrm{L}$ \cite{Bender2002b}. However, this requires $\lsym$ to be unitary; only then $\op{A}$ is antiunitary and thus an actual symmetry.

Analogously to \cref{eq:left_symmetrisation} we define the right symmetrisation operator $\rsym$ satisfying
\begin{equation}
  \op{H} \rsym = \rsym \adjg{\op{H}}
  \label{eq:right_symmetrisation}
\end{equation}
via the left-hand eigenvalue equation
\begin{equation}
  \adjg{\op{H}} \lket{\psi_n} = \conjg{E}_n \lket{\psi_n}
  \period
  \label{eq:left_eigenvalue_eq_right}
\end{equation}
The operator $\rsym$ transforms left eigenstates of $\op{H}$ into right eigenstates of $\op{H}$, i.\,e.\ $\rsym \lket*{\psi_n} = \rket*{\tpsi_n}$, thus complementing $\lsym$. As before, we can define an antilinear operator
\begin{equation}
  \op{T}_\mathrm{R} = \sum_m \rketbra{\psi_m}{\op{K}}{\psi_m}
  \label{eq:right_antiunitary_op}
\end{equation}
satisfying $\op{H} \op{T}_\mathrm{R} = \op{T}_\mathrm{R} \adjg{\op{H}}$,
so that
\begin{equation}
  \rsym = \op{A} \op{T}_\mathrm{R}
  \label{eq:unitary_right_symmetrisation_op}
\end{equation}
satisfies \cref{eq:right_symmetrisation}.

Further, the combination $\rsym \lsym$ commutes with $\op{H}$ by definition,
\begin{equation}
  \rsym \lsym \op{H}
    = \rsym \adjg{\op{H}} \lsym
    = \op{H} \rsym \lsym
  \period
  \label{eq:right_left_quasi-Hermiticity}
\end{equation}
Therefore, $\rsym \lsym$ and $\op{H}$ share the same eigenbasis. Thus, $\rsym \lsym$ is diagonal with respect to the bi-orthogonal product
\begin{equation}
  \left( \psi_n \middle\vert \psi_m \right)
    = \lrbraket{\psi_n}{\psi_m}
    = \delta_{nm}
  \label{eq:bi-orthogonal_product}
\end{equation}
defined in \cite{Moiseyev2011}. With a suitable choice of the symmetrisation operators the eigenvalues of $\rsym \lsym$ are unity. If $\lsym$ is invertible, then a Hamiltonian is, if at all, always symmetrisable from both sides due to
\begin{equation}
  \rsym \propto \inv{\lsym}
  \period
\end{equation}

For finite-dimensional Hilbert spaces there exists another important property: With the cyclic property of the determinant one can easily show that the coefficients of the characteristic polynomial must be real \cite{Bender2019}. Because of the fundamental theorem of algebra the spectrum consists only of complex-conjugated pairs. In contrast to \cref{eq:left_symmetrisation}, which is only a necessary condition \cite{Nixon2016a}, the reality of the coefficients of the characteristic polynomial is a sufficient condition to obtain a complex-conjugated spectrum \cite{Bender2010}, as it ensures that the kernel of the symmetrisation operators are empty. We return to this discussion in \cref{sub:semi_symmetrisation} and make use of this property in \cref{sec:applications}.

An operator satisfying \cref{eq:right_left_quasi-Hermiticity} is called either quasi-Hermitian \cite{Dieudonne1961,Scholtz1992}, $\lsym$-Hermitian \cite{Langer1962,Kuehne1964,Pease1965}, pseudo-Hermitian \cite{Froissart1959,Lee1969,Mostafazadeh2002a,Mostafazadeh2002b,Mostafazadeh2002c,Albeverio2004}, crypto-Hermitian \cite{Feinberg1999,Smilga2008a,Smilga2008b}, or generalised Hermitian \cite{Sun2010}, depending on the properties of $\lsym$. However, in contrast to these works the concept of symmetrisation considered here does not require the symmetrisation operators to be necessarily invertible \cite{Siegl2009,Nixon2016a} or their kernels to be empty.

We want to emphasise that we do not understand symmetrisation as a generalisation or extension of \ac{QM} as in \cite{Bender2002a,Mostafazadeh2010}, where the inner product is redefined by \cref{eq:bi-orthogonal_product}. This is possible, because in non-Hermitian \ac{QM} the left and right energy eigenstates form a bi-orthogonal basis. While left and right eigenstates coincide in Hermitian \ac{QM}, so that right and left eigenstates are orthogonal, respectively, in non-Hermitian \ac{QM} neither the left nor the right eigenstates can form a basis. Therefore, the Hermitian inner product of two right or two left states cannot be interpreted in the same manner as in Hermitian \ac{QM}. There exist two possible approaches to deal with this, which correspond to different interpretations:
\begin{enumerate}
  \item The bi-orthogonal product \labelcref{eq:bi-orthogonal_product} can be used as an inner product, which preserves some of the mathematical properties of \ac{QM}. However, by introducing a metric via $\lsym$, a new quantum theory is defined in which the left and right eigenstates of the eigenvalues of the Hamiltonian form a bi-orthogonal basis. By doing so, one defines a new Hilbert space in which Hamiltonians satisfying \cref{eq:left_symmetrisation} are effectively Hermitian. The motivation for the bi-orthogonal product is mainly of mathematical nature, as it allows for an elegant mathematical description of non-Hermitian \ac{QM}, similar to the usual Hermitian \ac{QM}. In this context the operators $\lsym$ and $\rsym$ are often referred to as metric operators.

  \item The Hermitian inner product of \ac{QM} between left and right states, respectively, is still valid if the non-Hermiticity is considered as a pure consequence of boundary conditions, i.\,e.\ they effectively describe open quantum systems with non-unitary evolution (see e.\,g.\ the concluding remarks in \cite{Jones2008}). Such Hamiltonians can, for example, be derived by using projection-operator methods \cite{Feshbach1958,Feshbach1962}. They are effectively described by a complex potential \cite{Sternheim1972}, so that the resulting Hamiltonian is effectively non-Hermitian and can be written in a complex-symmetric form \cite{Santra2002}.
\end{enumerate}
Both of these choices for the inner product can be seen as analytical continuations of the inner product of Hermitian \ac{QM} into the non-Hermitian domain.

In the following we understand non-Hermitian \ac{QM} as an effective description of open quantum systems. Hence, we must use the Hermitian inner product of \ac{QM} to obtain physical quantities. However, in this interpretation the left and right states are only parts of some larger orthogonal states describing a Hermitian system containing also the environment; thus, they do not form a basis on their own, respectively. \Cref{eq:left_symmetrisation,eq:right_symmetrisation} are then considered as properties of a given open quantum system, where $\lsym$ and $\rsym$ are constructed and utilised specifically for the given Hamiltonian \cite{Scholtz1992}.

The only exception has to be made for exceptional points, at which a certain number of energies and their corresponding left- and right-hand states coincide simultaneously and become self-orthogonal \cite{Moiseyev2011}. Hence, for numerical calculations in the vicinity of an exceptional points the states must be normalised using the bi-orthogonal product.

Suitable choices for the symmetrisation operators are given by \cite{Mostafazadeh2002d,Solombrino2002}
\begin{subequations}
  \begin{align}
    \lsym &= \sum_n p_n^{\rmL[0]} \lketbra*{\psi_n^0}{\psi_n^0}
      + \sum_m \qty[p_m^{\rmL[+]} \lketbra*{\psi_m^-}{\psi_m^+}
      + p_m^{\rmL[-]} \lketbra*{\psi_m^+}{\psi_m^-}]
      \comma
    \label{eq:left_symmetrisation_op_general}
    \\
    \rsym &= \sum_n p_n^{\rmR[0]} \rketbra*{\psi_n^0}{\psi_n^0}
      + \sum_m \qty[p_m^{\rmR[+]} \rketbra*{\psi_m^-}{\psi_m^+}
      + p_m^{\rmR[-]} \rketbra*{\psi_m^+}{\psi_m^-}]
    \label{eq:right_symmetrisation_op_general}
  \end{align}
\end{subequations}
with constant coefficients $p_k^q$, where $n$ runs over all real energies ($0$) and $m$ runs over all complex-conjugate pairs of energies with positive ($+$) and negative ($-$) imaginary parts. The terms corresponding to the complex part of the spectrum are traceless; hence, $\lsym$ and $\rsym$ are indefinite. For this reason there occur states of the Hamiltonian with negative norm, which must to be considered unphysical \cite{Scholtz1992} and may be excluded by superselection rules \cite{Brody2017}. Nevertheless, the occurrence of symmetric pairs of complex eigenvalues can be considered physical in general, as these can be understood as emission and absorption phenomena \cite{Jakob2003}.

The concept of indefinite metrics is not new both in physics \cite{Dirac1942,Pauli1943} and mathematics \cite{Pontryagin1944,Krein1950}. Such an indefinite metric gives rise to a Krein space \cite{Langer1962,Azizov1979}, in which, for example, non-Hermitian operators with real spectrum are self-adjoint.

\subsection{Semi-symmetrisation}%
\label{sub:semi_symmetrisation}

In this section we consider the case in which the symmetrisation operators are not invertible. However, as it was not directly required for them to be invertible in the first place, symmetrisation of $\op{H}$ is still possible to a certain degree.

A similar concept was described in \cite{Nixon2016a}, however, with the distinction that we do not require the symmetrisation operators to be differential operators. Therefore, semi-symmetrisation does not require the specific type of potential proposed by Wadati in \cite{Wadati2008}, which is not physical in all types of systems. To make this clear, a discussion on different types of complex potentials follows in \cref{sub:physical_complex_potentials}.

Consider a left symmetrisation operator $\lsym$ with $\det\lsym = 0$. Clearly, there is no inverse of $\lsym$ and its kernel is non-empty. Nevertheless, it is still possible to satisfy \cref{eq:left_symmetrisation} if we demand that the elements in the kernel of $\lsym$ are exclusively right eigenstates $\rket{\varphi_n}$ of $\op{H}$. For these states \cref{eq:left_eigenvalue_eq_lsym} is fulfilled trivially as $\lsym \rket{\varphi_n} = 0$. However, the corresponding eigenvalues are neither real nor part of a conjugate pair; these energies form isolated complex resonances in the spectrum of $\op{H}$.

Since \cref{eq:right_left_quasi-Hermiticity} still remains valid, $\rsym \lsym$ can be considered as an identity with respect to the right eigenstates of $\op{H}$ which are not in the kernel of $\lsym$ and acts as an annihilation operator for states from the kernel of the left symmetrisation operator. Though $\rsym$ is not the inverse of $\lsym$, they are semi-inverse. Following the definition $A B A = A$ \cite{Frame1964}, where $B$ is the semi-inverse of $A$, we find
\begin{equation}
  \lsym \rsym \lsym \rket*{\psi_n}
    = \lsym \rsym \lket*{\tpsi_n}
    = \lsym \rket*{\psi_n}
  \period
\end{equation}
Hence we call such a system semi-symmetrisable.

Semi-symmetrised operators possess the properties of symmetrised operators on a subspace spanned by eigenstates of the Hamiltonian which are not in the kernels of the symmetrisation operators. The dimension of this subspace is given by the rank of $\lsym$. With respect to the definition \labelcref{eq:left_symmetrisation_op_general}, it is obvious that, because of bi-orthogonality, for any left eigenstate we exclude from the sums, the corresponding right eigenstate is part of the kernel of the symmetrisation operator. The same applies to the definition \labelcref{eq:right_symmetrisation_op_general} of the right symmetrisation operator. Since the rank of the symmetrisation operators can take any value between 1 and the full rank, we conclude that any operator with a discrete number of real eigenvalues is semi-symmetrised, even if the spectrum is not entirely discrete. For this reason one can apply symmetrisation to a specific subspace, for example obtained by means of some approximation, without the need to care for the entirety of the spectrum. A specific example for a physical system which is solely semi-symmetrisable is the two-mode system with arbitrary gain and loss, for which a quantitative study can be found in \cref{sub:two_mode_systems_with_arbitrary_gain_and_loss}.

Following the line of argument in \cite{Dast2013b}, symmetrisation is in principle applicable to non-linear quantum systems, if the non-linear part also satisfies \cref{eq:left_symmetrisation,eq:right_symmetrisation}. Like for $\op{PT}$ symmetry, this might require that the wave function is invariant under the symmetrisation operators. However, in non-linear cases also the properties of linear algebra are lost. This makes an analytical treatment of non-linear, non-Hermitian quantum systems quite difficult, even for simple discrete systems. Hence, in most of these cases one must resort to numerical investigations, as done in \cite{Lunt2017}. Moreover, there exists no bi-orthogonality in non-linear non-Hermitian systems, which implies that every state defines its own pair of symmetrisation operators with \enquote{rank} 1, though the term is no longer well defined. For these reasons we only consider linear symmetrised quantum systems for now.

\subsection{Physical complex potentials}%
\label{sub:physical_complex_potentials}

In the previous sections multiple concepts to obtain physical spectra in the context of \cite{Jakob2003} (i.\,e.\ complex-conjugated) in non-Hermitian quantum systems were mentioned and discussed. Apart from the mathematical considerations, it is important to consider the types of potentials which these concepts allow. For example, $\op{PT}$ symmetry allows for arbitrary real and imaginary potentials, as long as they are symmetric and antisymmetric, respectively \cite{Bender1999}. This makes the concept of $\op{PT}$ symmetry applicable to almost every kind of physical system. Nevertheless, the requirement of exact symmetry can be demanding, especially in the presence of perturbations in experiments.

In contrast to other approaches we intend to construct potentials providing real and complex-conjugated spectra for some given physical conditions like a given gain and loss distribution. In most other approaches Wadati-type potentials are used \cite{Nixon2016a,Nixon2016b,Yang2017,Rivet2018}. They allow only for a specific type of complex potentials of the form
\begin{equation}
  V(x) = w^2 (x) + \imag \pdv{w}{x} (x)
  \comma
  \label{eq:Wadati_potential}
\end{equation}
where $w(x)$ is an arbitrary real function defining the whole system. As mentioned above, such a potential does, however, not yield a physical interpretation in every physical system. An example of a system with a physical interpretation of such potentials is given by the pressure waves in \cite{Rivet2018}. However, in case of Bose-Einstein condensates which are localised in distinct potential wells as considered in \cite{Lunt2017}, a differential imaginary potentials cannot be interpreted directly. This is, because the imaginary part of the potential describes in- and out-coupling of particles. Therefore, a Wadati-type potential would always describe in-coupling and out-coupling of particles in the same well. It is thus unclear how to extract the overall gain and loss of particles in each well, respectively. Moreover, such quantum systems can hardly be realised (cf.\ \cite{Altinisik2019} and references therein).

We now want to focus on open quantum systems described by Hamiltonians with a complex multi-well potential, which describes non-symmetric gain and loss in each well, respectively. For its description we have to rely on the concept of symmetrisation, as it allows for asymmetric potentials while not requiring the specific potential \labelcref{eq:Wadati_potential}. The Hamiltonians of such systems are complex symmetric, i.\,e.\ $\transp{\op{H}} = \op{H}$, thus being real apart from its diagonal. This involves no loss of generality, since any matrix can be transformed into a complex-symmetric form \cite{Santra2002}. For complex-symmetric Hamiltonians the set of bi-orthogonal states is complete \cite{Naimark1968,Sternheim1972}. For any right eigenstate of $\op{H}$ the corresponding left eigenstate is then easy to find: conjugating \cref{eq:left_eigenvalue_eq_right} yields
\begin{equation}
  \transp{\op{H}} \lket*{\conjg{\tpsi}_n}
    = \op{H} \lket*{\conjg{\tpsi}_n}
    = \conjg{E}_n \lket*{\conjg{\tpsi}_n}
  \period
\end{equation}
It is immediately clear that left and right eigenstates of a complex-symmetric Hamiltonian are complex conjugate with respect to their bra or ket forms, respectively. Note that the left bra and right ket eigenstates for complex-symmetric Hamiltonians are transpose of each other, because they share the symmetry of the Hamiltonian.

An interesting property arises in the context of complex potentials. If one considers \cref{eq:left_symmetrisation} as a property of a given system with an imaginary potential, then \cref{eq:right_symmetrisation} is a property of the same system with the negative imaginary potential. In other words, if the system is symmetrisable from both sides simultaneously, then there exist two suitable gain and loss distributions, each corresponding to one symmetrisation operator. The inversion of gain and loss corresponds to time reversal, so that
\begin{equation}
  \op{H} = \op{T} \adjg{\op{H}} \op{T}
\end{equation}
with the involutory time-reversal operator $\op{T}$. By plugging this into \cref{eq:left_symmetrisation} and comparing the result to \cref{eq:right_symmetrisation} we find
\begin{equation}
  \rsym = \op{T} \lsym \op{T}
  \period
\end{equation}
This is a consequence of the dual character of $\op{H}$ and $\adjg{\op{H}}$, which is in case of complex-symmetric Hamiltonians defined with respect to the complex conjugate rather than the Hermitian adjoint. An interesting discussion on this duality can be found in \cite{Simon2018}, where it is shown that left eigenstates evolve backwards in time compared to right eigenstates and vice versa.

\section{Application to matrix models}%
\label{sec:applications}

\begin{figure}[b]
  \centering
  \includegraphics[width=0.9\columnwidth]{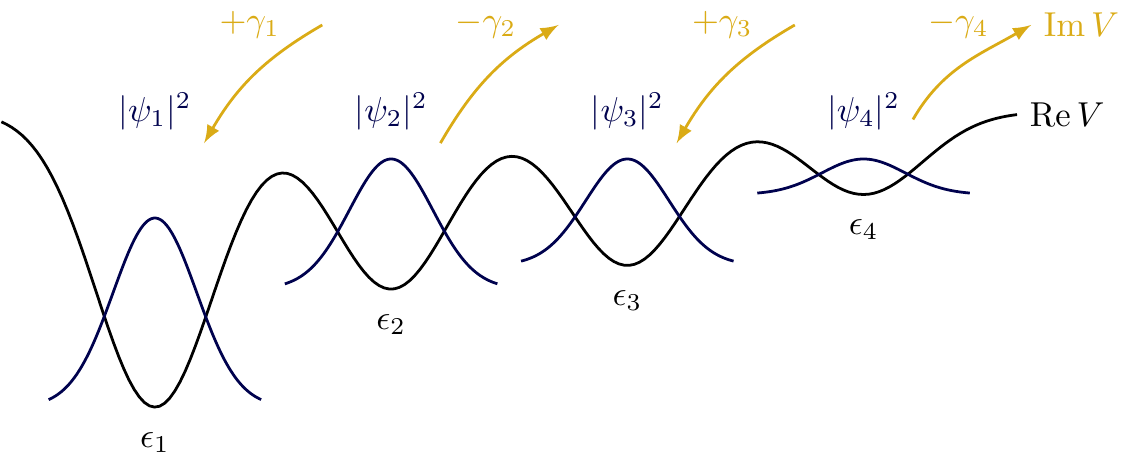}
  \caption{Complex one-dimensional multi-well potential with localised gain and loss. The wave function is localised inside the wells and can be characterised by the occupation numbers $\abs{\psi_k}^2$. The real and imaginary parts of the potential are described by the parameters $\epsilon_k$ (on-site energies) and $\gamma_k$ (coupling strengths) in each well, respectively.}%
  \label{fig:fig1}%
\end{figure}

In this section we apply the concept of symmetrisation discussed in \cref{sec:theory} to complex multi-well potentials, where the imaginary part describes in- and out-coupling of particles in each well as shown in \cref{fig:fig1}. By assuming that the potential wells are deep and the energies are well separated such systems can be described by a matrix model with a complex potential (see e.\,g.\ \cite{Kreibich2014}). This specifically means that we deal with finite-dimensional Hilbert spaces in contrast to \cite{Altinisik2019}, where the symmetrisation of the spatially extended potentials is considered.

\subsection{Two-mode systems with arbitrary gain and loss}%
\label{sub:two_mode_systems_with_arbitrary_gain_and_loss}

We start by investigating the formation of steady states in an open linear two-mode system described by a linear non-Hermitian matrix model. Two-mode systems are well studied both for $\op{PT}$-symmetric potentials \cite{Graefe2008a,Guo2009,Rueter2010,Cartarius2012,Graefe2012,Dast2013a} and recently also for arbitrary gain and loss potentials \cite{Ding2016,Liu2017}. The most general symmetrisation operator for two-dimensional systems is also already known \cite{Mostafazadeh2004,Mostafazadeh2006}. In the following we discuss the two-mode system with respect to symmetrisation.

The Hamiltonian $\op{H}$ of the system reads in atomic units
\begin{equation}
  \op{H} = \mqty(
    \epsilon_1 + \imag \gamma_1 & -J \\
    -J & \epsilon_2 + \imag \gamma_2
  )
  \comma
  \label{eq:Hamiltonian_2}
\end{equation}
where $J > 0$ is the transition rate between the two sites and the real potential is given by the on-site energies $\epsilon_1$ and $\epsilon_2$. The imaginary part of the potential makes the Hamiltonian non-Hermitian and allows for the description of an effectively open system via the in- and out-coupling parameters $\gamma_1$ and $\gamma_2$. Note that the tunnelling rate could without loss of generality be set to $J = 1$, which corresponds to a specific choice of the energy scale. However, we keep the parameter in the following, so that the reader may keep track of where the off-diagonal elements enter into calculations.

The reality of the characteristic polynomial of the Hamiltonian \labelcref{eq:Hamiltonian_2} allows for checking efficiently whether the Hamiltonian is symmetrised or not. This evaluation yields the two conditions
\begin{subequations}
  \begin{align}
    \gamma_1 + \gamma_2 &= 0
      \comma
      \label{eq:2_mode_linear_coef}
    \\
    \epsilon_1 \gamma_2 + \gamma_1 \epsilon_2 &= 0
      \label{eq:2_mode_constant_coef}
    \period
  \end{align}
\end{subequations}
Of these \cref{eq:2_mode_linear_coef} is only satisfied for an antisymmetric imaginary potential and thus by a $\op{PT}$-symmetric Hamiltonian. Since the reality of the characteristic polynomial is a sufficient condition for a real or complex-conjugate spectrum, we can conclude that there are no asymmetric potentials with entirely real spectrum. The symmetrisation condition \labelcref{eq:left_symmetrisation}, in contrast, is not a sufficient condition and thus it might be possible to obtain a system with at least some real eigenvalues. Therefore, we continue investigating the symmetrisation or semi-symmetrisation of the Hamiltonian \labelcref{eq:Hamiltonian_2}.

In two-dimensional spaces we may conveniently write the symmetrisation operator in terms of Pauli matrices,
\begin{equation}
  \lsym = \sum_{n = 0}^3 \lsym[n] \sigma_n
    \comma
    \label{eq:lsym_2}
\end{equation}
where $\sigma_0 = \id$ and $\lsym[n] \in \realnumbers$, so that $\lsym$ is Hermitian.

With \cref{eq:Hamiltonian_2,eq:lsym_2} the symmetrisation condition \labelcref{eq:left_symmetrisation} yields
\begin{equation}
  \mqty(
    \gamma_1 + \gamma_2 & 0 & 0 & \gamma_1 - \gamma_2 \\
    0 & \gamma_1 + \gamma_2 & \epsilon_1 - \epsilon_2 & 0 \\
    0 & -\qty(\epsilon_1 - \epsilon_2) & \gamma_1 + \gamma_2 & -2 J \\
    \gamma_1 - \gamma_2 & 0 & 2 J & \gamma_1 + \gamma_2
  ) \mqty( \lsym[0] \\ \lsym[1] \\ \lsym[2] \\ \lsym[3] )
    = 0
    \period
    \label{eq:coefficient_matrix}
\end{equation}
A solution for this equation only exists if the determinant of the coefficient matrix vanishes, i.\,e. if
\begin{equation}
  \qty(\gamma_1 + \gamma_2)^2
    \qty[\qty(\gamma_1 + \gamma_2)^2 - \qty(\gamma_1 - \gamma_2)^2 + 4 J^2]
  + \qty(\epsilon_1 - \epsilon_2)^2
    \qty[\qty(\gamma_1 + \gamma_2)^2 - (\gamma_1 - \gamma_2)^2]
  = 0
  \period
    \label{eq:determinant_coefficnet_matrix}
\end{equation}

For a symmetric potential with $\epsilon_1 = \epsilon_2$ and $\gamma_1 = - \gamma_2 = \gamma$ both terms in \cref{eq:determinant_coefficnet_matrix} vanish identically. The solution of \cref{eq:coefficient_matrix} then is
\begin{equation}
  \lsym = \mqty(
    \lsym[0] & \lsym[1] - \imag \frac{\gamma}{J} \lsym[0] \\
    \lsym[1] + \imag \frac{\gamma}{J} \lsym[0] & \lsym[0]
  )
    \comma
  \label{eq:lsym_2_symmetric}
\end{equation}
which has the two degrees of freedom $\lsym[0]$ and $\lsym[1]$. We can obtain the corresponding right symmetrisation operator by simply changing the sign of $\gamma$, which corresponds to the adjoint Hamiltonian; hence, $\rsym = \conjg{\lsym}$. For $\lsym[0] = 0$ and $\lsym[1] = 1$ we retrieve the parity operator $\op{P}$.

For $J \neq 0$ there is no other choice of parameters for which the first term in \cref{eq:determinant_coefficnet_matrix} vanishes. However, by choosing
\begin{equation}
  \epsilon_1 - \epsilon_2
    = \pm \imag \qty(\gamma_1 + \gamma_2)
    \sqrt{1 + \frac{J^2}{\gamma_1 \gamma_2}}
    \label{eq:difference_onsite_energies}
\end{equation}
with $-J^2 < \gamma_1 \gamma_2 < 0$, the condition \cref{eq:determinant_coefficnet_matrix} is satisfied. The lower bound of the product $\gamma_1 \gamma_2$ stems from the fact that if we assume that only either $\gamma_1$ or $\gamma_2$ is nonzero, then \cref{eq:determinant_coefficnet_matrix} can only be satisfied if the modes are decoupled, i.\,e.\ $J = 0$, or if there is no gain or loss at all, which contradicts the assumption. The upper bound is required for the expression under the square root to be negative, so that the difference of the on-site energies is a real quantity. Hence, a symmetrisable operator of the form \labelcref{eq:Hamiltonian_2} always requires the presence of both gain and loss. This appears to be intuitive, in particular if we recall that real energies describe stationary states in open quantum systems. However, as shown in \cite{Nixon2016a}, there exist potentials in unbound systems which do not require gain at all, which is hard to grasp from a physical point of view.

From a mathematical point of view, there is an infinite number of systems which can fulfil the relation \labelcref{eq:difference_onsite_energies}, as we only require a specific difference in the on-site energies. Since we are free, however, to gauge the energy scale, we only have to consider two physically different situations which correspond to a system $\op{H}$ with parameters $(\epsilon_1, \epsilon_2)$ according to \cref{eq:difference_onsite_energies} and its time-reversed counterpart $\adjg{\op{H}}$. We suppose a system to evolve forward in time if the probability or particle current in the system is directed from left to right (cf.\ \cref{fig:fig1}). Consequently, if the current is directed from right to left the system is evolving backwards in time.

Note that \cref{eq:difference_onsite_energies} corresponds to the results (5) found by Lunt \etal{} in \cite{Lunt2017} with the parametrisation $\epsilon_1 \equiv a_R$, $\epsilon_2 \equiv -a_R$, $\gamma_1 \equiv \gamma \qty(1 + a_I)$, and $\gamma_2 \equiv - \gamma \qty(1 - a_I)$ with $J = 1$. However, the left-hand side of their equation should be squared, which can be checked both by comparison with \cref{eq:difference_onsite_energies} and direct calculation.

By plugging \cref{eq:difference_onsite_energies} into \cref{eq:coefficient_matrix} we can calculate the symmetrisation operator, which in matrix form reads
\begin{equation}
  \lsym = \mqty(
    -\frac{J}{\gamma_1}
      & -\imag \qty(1 \pm \sqrt{1 + \frac{J^2}{\gamma_1 \gamma_2}})
    \\
    \imag \qty(1 \mp \sqrt{1 + \frac{J^2}{\gamma_1 \gamma_2}})
      & \frac{J}{\gamma_2}
  ) \lsym[2]
\end{equation}
and has only the one degree of freedom $\lsym[2]$ in contrast to \cref{eq:lsym_2_symmetric}. This matrix is non-invertible as $\det \lsym = 0$ independently of the choice of $\lsym[2]$. Hence, $\rank \lsym = 1$, which immediately implies that there exists one real eigenvalue $\mu$ of the Hamiltonian \labelcref{eq:Hamiltonian_2}. By suitably shifting the energy scale this real eigenvalue can take the form $\mu = 0$ and its existence is determined by $\det \op{H} = 0$. The solutions of this condition are given by \cref{eq:difference_onsite_energies}. We can thus state the situation as follows: there are two physically different systems with $\epsilon_1 > \epsilon_2$ and $\epsilon_1 < \epsilon_2$, respectively, which satisfy \cref{eq:difference_onsite_energies}. For both these systems there exist two imaginary potentials $\gamma_1 > \gamma_2$ (forward in time) and $\gamma_1 < \gamma_2$ (backwards in time), respectively, so that each system has one real eigenvalue. Semi-symmetrisation thus connects four Hamiltonians $\op{H}_+$, $\adjg{\op{H}}_+$, $\op{H}_-$, and $\adjg{\op{H}}_-$, where the index refers to the sign in \cref{eq:difference_onsite_energies}.

\begin{figure}
  \includegraphics[width=\textwidth]{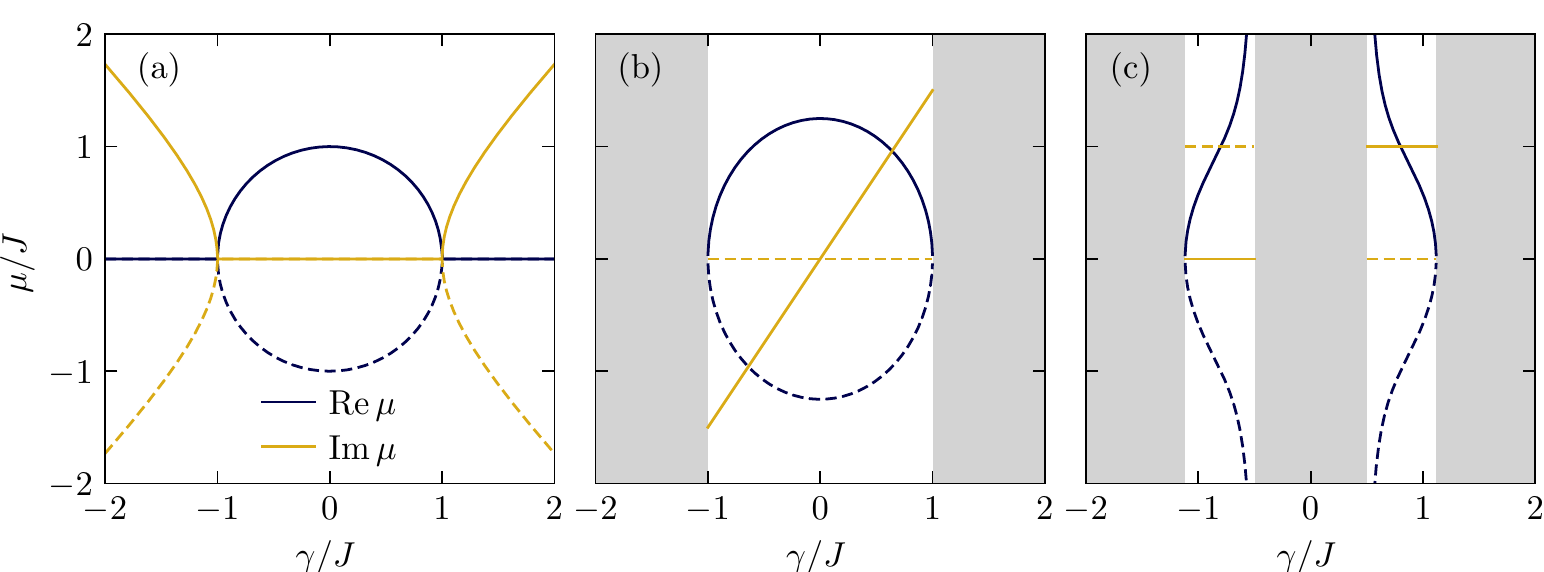}
  \caption{Real and imaginary parts of the eigenvalues \labelcref{eq:mu_2} for \cref{eq:difference_onsite_energies} with $\epsilon_1 \ge \epsilon_2$ and the different parametrisations (a) $\gamma_1 = -\gamma_2 = \gamma$, (b) $\gamma_1 = 2 \gamma$, $\gamma_2 = -\flatfrac{\gamma}{2}$, and (c) $\gamma_1 = \gamma + \flatfrac{1}{2}$, $\gamma_2 = -\gamma + \flatfrac{1}{2}$. The solutions $\mu^+$ and $\mu^-$ are drawn as solid and dashed lines, respectively. For antisymmetric gain and loss there are two real solutions for $\abs{\gamma} < J$, which form a conjugate pair for $\abs{\gamma} > J$. For asymmetric gain and loss there exists one real and one complex solution, respectively, if the Hamiltonian is symmetrised. The grey shaded areas indicate the region where no symmetrisation is possible.}%
  \label{fig:fig2}%
\end{figure}

\begin{figure}
  \includegraphics[width=\columnwidth]{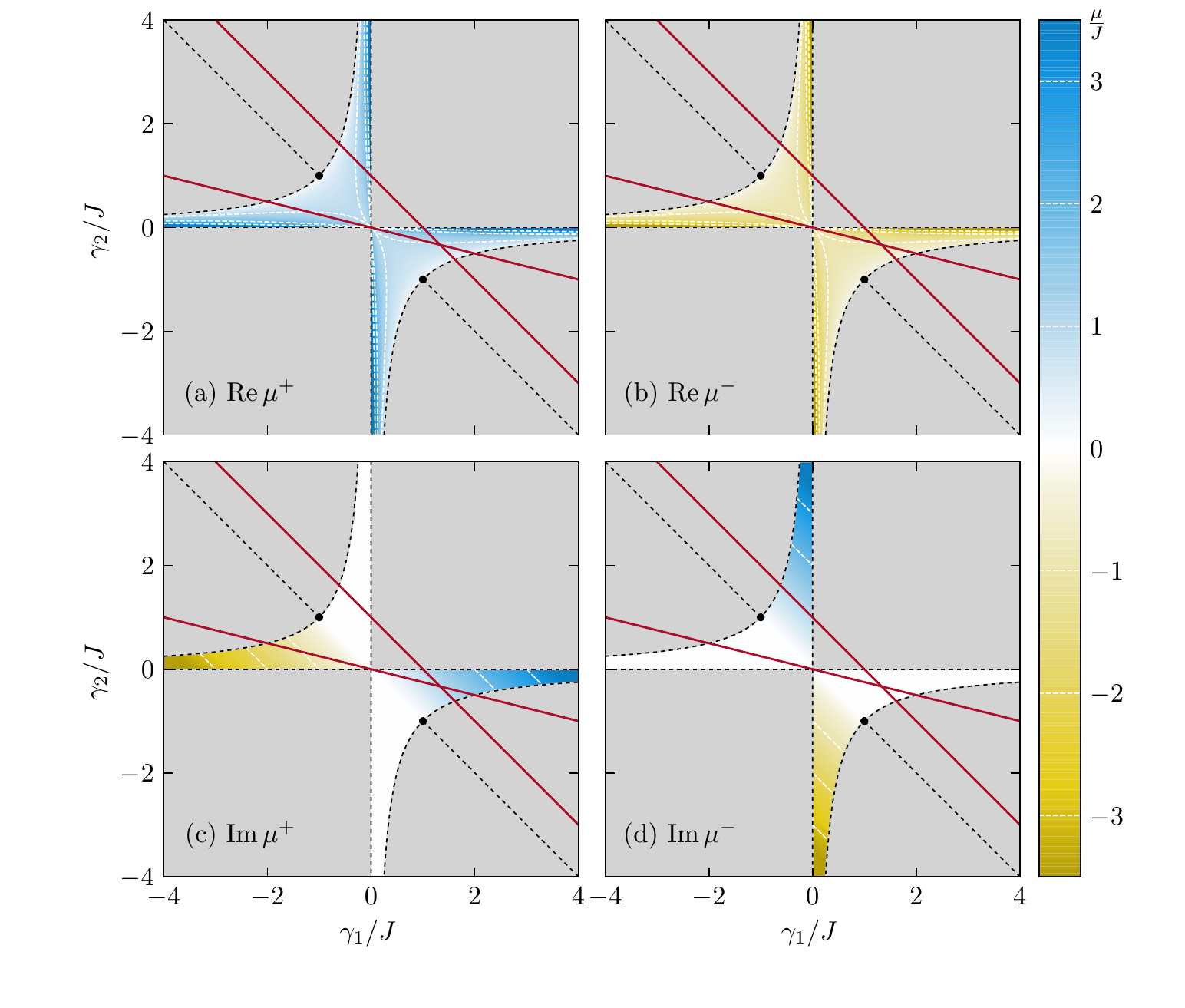}
  \caption{Real and imaginary parts of the eigenvalues \labelcref{eq:mu_2} as functions of $\gamma_1$ and $\gamma_2$, where $\epsilon_1 > \epsilon_2$ are determined by \cref{eq:difference_onsite_energies}. The two solid straight lines show the $\gamma$ axes for the parametrisations in \csubref{fig:fig2}{b} and \csubref{fig:fig2}{c}. The black dashed lines indicate the borders between the regions where the system is symmetrised and where it is not. The white dashed lines are contour lines and the exceptional points are marked by dots. The case $\epsilon_1 < \epsilon_2$ looks identical but with (c) and (d) exchanged.}%
  \label{fig:fig3}%
\end{figure}

The general solutions for the eigenvalue of the two-mode system are given by
\begin{equation}
    \mu = \frac{1}{2} \qty[
        \qty(\epsilon_1 + \epsilon_2) + \imag \qty(\gamma_1 + \gamma_2)
        \pm \sqrt{
          \qty[\qty(\epsilon_1 - \epsilon_2) + \imag \qty(\gamma_1 - \gamma_2)]^2
        + 4 J^2
        } ]
    \period
  \label{eq:mu_2}
\end{equation}
These solutions $\mu^+$ and $\mu^-$ are plotted in \cref{fig:fig2} for different parametrisations of gain and loss. \Csubref{fig:fig2}{a} shows the case with a symmetric potential, i.\,e.\ $\gamma_1 = -\gamma_2 = \gamma$, which is well known from $\op{PT}$ symmetry and will not be discussed in the following; instead we refer to \cite{Graefe2012} for a detailed discussion on the $\op{PT}$-symmetric two-mode system.

\Csubref{fig:fig2}{b} shows a system with asymmetric gain and loss
\begin{equation}
  \gamma_1 = 2 \gamma
    \comma
  \quad \gamma_2 = -\frac{\gamma}{2}
    \period
  \label{eq:parametrisation_gain_loss_rotated}
\end{equation}
The on-site energies are chosen symmetrically as
\begin{equation}
  \epsilon_1 = -\epsilon_2 = \frac{\Delta \epsilon}{2}
    \comma
  \label{eq:parametrisation_onsite_energies}
\end{equation}
where $\Delta \epsilon$ corresponds to \cref{eq:difference_onsite_energies}. Thus, $\epsilon_1 + \epsilon_2 = 0$ in \cref{eq:mu_2}. We use the freedom in the sign of $\Delta \epsilon$ and choose $\epsilon_1 > \epsilon_2$. For $\abs{\gamma} < J$ the spectrum resembles the symmetric case in \csubref{fig:fig2}{a} but with the imaginary part of $\mu^+$ rotated about the origin $\gamma = 0$, where the system is Hermitian, while $\mu^-$ remains entirely real. The real parts of $\mu^+$ and $\mu^-$ are still symmetric but the ratio of the axes changes by a factor of $1.25$. In the domain $\abs{\gamma} > J$ there exist no real solutions at all, since \cref{eq:difference_onsite_energies} becomes imaginary and cannot be used to determine the on-site energies anymore. At $\gamma = \pm J$ no bifurcations occur in comparison to the symmetric case, as the imaginary parts of $\mu^+$ are nonzero.

In \csubref{fig:fig2}{c} another parametrisation of asymmetric gain and loss is shown with
\begin{equation}
  \gamma_1 = \gamma + \frac{1}{2}
    \comma
  \quad
  \gamma_2 = -\gamma + \frac{1}{2}
    \period
  \label{eq:parametrisation_gain_loss_shifted}
\end{equation}
and \cref{eq:parametrisation_onsite_energies}, so that $\gamma_1 + \gamma_2 = 1$. We again find a deformation of the solutions of the symmetric case in \csubref{fig:fig2}{a}. However, there are no solutions on the interval $\qty[-\flatfrac{1}{2}, \flatfrac{1}{2}]$, where the gain and loss parameters \labelcref{eq:parametrisation_gain_loss_shifted} have the same sign; i.\,e.\ for $-\flatfrac{1}{2} < \gamma < 0$ there is only gain and for $0 < \gamma < \flatfrac{1}{2}$ there is only loss. For $\abs{\gamma} > J$ there are no solutions either, for the same reasons discussed above. In the remaining parameter regions there are solutions with either $\Im \mu^+ = 0$ ($\gamma < 0$) or $\Im \mu^- = 0$ ($\gamma > 0$). The offset of the respective other imaginary parts are given by $\imag \qty(\gamma_1 + \gamma_2)$, which is zero in the symmetric case. We further find that the real parts of the energy eigenvalues diverge at $\gamma = \pm \flatfrac{1}{2}$. By approaching the regions where only either gain or loss occurs the pairs $\qty(\gamma_1, \gamma_2)$ become increasingly unsuitable for sustaining the properties we want to impose onto the system. This is reflected by the divergence of the term \labelcref{eq:difference_onsite_energies} at $\gamma = \pm \flatfrac{1}{2}$ due to
\begin{equation}
  \gamma_1 \gamma_2 = \frac{1}{4} - \gamma^2
\end{equation}
and the subsequent divergence of $\Re \mu^+$ and $\Re \mu^-$.

\Cref{fig:fig3} shows the real and imaginary parts of the energy eigenvalues \labelcref{eq:mu_2} as functions of $\qty(\gamma_1, \gamma_2)$. The area where real solutions of \cref{eq:difference_onsite_energies} exist is enclosed by hyperbolas and the axes in the second and forth quadrants. In the first and third quadrants, there exists no solution at all, as they correspond to pure gain or loss. For $\gamma_1 = -\gamma_2$, which corresponds to the off-diagonal, both $\mu^+$ and $\mu^-$ are real and $\Delta \epsilon = 0$, which corresponds to the case of a symmetric potential.

The slices in the $(\gamma_1, \gamma_2)$ space corresponding to the parametrisations \eqref{eq:parametrisation_gain_loss_rotated} and \eqref{eq:parametrisation_gain_loss_shifted} are shown by the two solid, straight lines. \Csubrefs{fig:fig3}{c} and (d) show that the imaginary parts of $\mu^+$ and $\mu^-$ are growing linearly along directions parallel to the diagonal. Hence we find a constantly shifted imaginary part by slicing parallel to the off-diagonal and a, as it were, rotated imaginary part by slicing along any other direction through the origin. Since both real parts diverge towards the axes, the energy eigenvalues may be arbitrarily large by rotating the slice around the origin. Note that this divergence is a consequence of the matrix approximation and is not observed in a continuous description \cite{Altinisik2019}.

To conclude, the findings of Lunt \etal{} are an example of a semi-symmetrised Bose-Einstein condensate. The discussions in \cite{Lunt2017} provide some further insight into the physical meaning of this mathematically rather abstract concept. Lunt \etal{} also discuss a two-mode system with asymmetric gain and loss but with a symmetric trapping potential, which corresponds to $\epsilon_1 = \epsilon_2 = 0$ and
\begin{equation}
  \gamma_1 = \pm J \sqrt{\frac{1 + a}{1 - a}}
    \comma
  \quad
  \gamma_2 = \mp J \sqrt{\frac{1 - a}{1 + a}}
    \comma
\end{equation}
where $a \in \realnumbers$ is a free parameter. This Hamiltonian also yields a single real eigenvalue, though it is not semi-symmetrisable because of \cref{eq:difference_onsite_energies}. Hence, semi-symmetrisation is neither necessary nor sufficient for the occurrence of real eigenvalues. However, in the double-well potential it allows for a systematic approach to find a class of real eigenvalues with special properties, namely that the corresponding states are stable if the Hamiltonian becomes non-linear \cite{Lunt2017}.

\subsection{Few-mode systems}%
\label{sub:few_mode_systems}

Although few-mode systems with up to four dimensions can, in principle, be treated analytically as described in \cref{sub:two_mode_systems_with_arbitrary_gain_and_loss}, the number of equations grows quadratically, which is not feasible. However, by demanding the reality of the characteristic polynomial, the number of equations grows only linearly and enables us to investigate a three-dimensional model analytically. Since this condition is sufficient, we deal with fully symmetrised systems at first.

The Hamiltonian of the three-mode system possesses the same structure as the Hamiltonian \labelcref{eq:Hamiltonian_2},
\begin{equation}
  \op{H} = \mqty(
    \epsilon_1 + \imag \gamma_1 & -J & 0 \\
    -J & \epsilon_2 + \imag \gamma_2 & -J \\
    0 & -J & \epsilon_3 + \imag \gamma_3
  )
  \period
  \label{eq:Hamiltonian_3}
\end{equation}
Evaluation of the imaginary parts of the coefficients of the characteristic polynomial yields
\begin{subequations}
  \begin{align}
    \gamma_1 + \gamma_2 + \gamma_3 &= 0
      \comma
      \label{eq:3_mode_quadratic_coef}
    \\
    \epsilon_1 \gamma_2 + \gamma_1 \epsilon_2
      + \epsilon_2 \gamma_3 + \gamma_2 \epsilon_3
      + \epsilon_1 \gamma_3 + \gamma_1 \epsilon_3
        &= 0
      \comma
      \label{eq:3_mode_linear_coef}
    \\
    \gamma_1 \epsilon_2 \epsilon_3
      + \epsilon_1 \gamma_2 \epsilon_3
      + \epsilon_1 \epsilon_2 \gamma_3
      + \gamma_1 \gamma_2 \gamma_3
      - J^2 \qty(\gamma_1 + \gamma_3)
        &= 0
      \period
      \label{eq:3_mode_constant_coef}
  \end{align}
\end{subequations}
\Cref{eq:3_mode_quadratic_coef} shows, as already stated in \cite{Simeonov2016}, that the sum of all gain and loss terms must vanish. This arises from the condition that the trace of the Hamiltonian must be real, which corresponds to highest-order non-trivial coefficients of the characteristic polynomial (cf.\ \cref{eq:2_mode_linear_coef}), and appears to be a necessary consequence of balanced gain and loss (e.\,g.\ see \cite{Nixon2016a}). However, as we showed in the two-dimensional case, there are stationary solutions with real energy even if \cref{eq:2_mode_linear_coef} is not satisfied. Hence, we emphasise that balanced gain and loss has to be considered per state, meaning that one also has to consider the occupation numbers $n_k$ in each well. Thus, balanced gain and loss means
\begin{equation}
  \sum_k n_k \gamma_k = 0
    \period
\end{equation}

It is easy to spot the trivial solutions of \cref{eq:3_mode_quadratic_coef,eq:3_mode_linear_coef,eq:3_mode_constant_coef}: the Hermitian potential, $\gamma_1 = \gamma_2 = \gamma_3 = 0$ and the symmetric potential, $\gamma_1 = -\gamma_3$, $\gamma_2 = 0$, $\epsilon_1 = \epsilon_3$, where $\epsilon_2$ can be chosen arbitrarily in the latter case. However, if either only $\gamma_1$ or $\gamma_3$ is zero, \cref{eq:3_mode_constant_coef} cannot be satisfied for $J \neq 0$. The other solutions of \cref{eq:3_mode_quadratic_coef,eq:3_mode_linear_coef,eq:3_mode_constant_coef} are given by
\begin{subequations}
  \begin{align}
    \gamma_1 &= -\qty(\epsilon_2 - \epsilon_3) \gamma_0
      \comma
    \label{eq:3_mode_solution_gamma_1}
    \\
    \gamma_2 &= \qty(\epsilon_1 - \epsilon_3) \gamma_0
      \comma
    \label{eq:3_mode_solution_gamma_2}
    \\
    \gamma_3 &= -\qty(\epsilon_1 - \epsilon_2) \gamma_0
      \comma
    \label{eq:3_mode_solution_gamma_3}
  \end{align}
  where
  \begin{equation}
    \gamma_0 = \pm \sqrt{
      \frac{\Delta \epsilon_{12}^3 + \Delta \epsilon_{23}^3
        - \Delta \epsilon_{13}^3 + 3 J^2 \Delta \epsilon_{13}}%
      {3 \Delta \epsilon_{12} \Delta \epsilon_{23} \Delta \epsilon_{13}}
    }
    \label{eq:3_mode_gamma_0}
  \end{equation}
  with $\Delta \epsilon_{kl} = \epsilon_k - \epsilon_l \neq 0$.
\end{subequations}

Of course the solutions \labelcref{eq:3_mode_solution_gamma_1,eq:3_mode_solution_gamma_2,eq:3_mode_solution_gamma_3} again only depend on the difference of the on-site energies; otherwise we would not be able to freely gauge the energy. Further, those solutions only exist if the term under the square root in \cref{eq:3_mode_gamma_0} is positive. By assuming that $\Delta \epsilon_{12} \Delta \epsilon_{23} > 0$, we find
\begin{equation}
  \Delta \epsilon_{12} \Delta \epsilon_{23} \le J^2
    \period
\end{equation}
However, for $\Delta \epsilon_{12} \Delta \epsilon_{23} < 0$, there are no solutions at all. Hence, $\Delta \epsilon_{12}$ and $\Delta \epsilon_{23}$ must have the same sign, which yields
\begin{equation}
  \epsilon_1 \lessgtr \epsilon_2 \lessgtr \epsilon_3
    \comma
\end{equation}
with two possible gain and loss distributions
\begin{equation}
  \gamma_1 \gtrless 0 \comma
  \quad \gamma_2 \lessgtr 0 \comma
  \quad \gamma_3 \gtrless 0
\end{equation}
in each case.

\begin{figure}
  \includegraphics[width=\columnwidth]{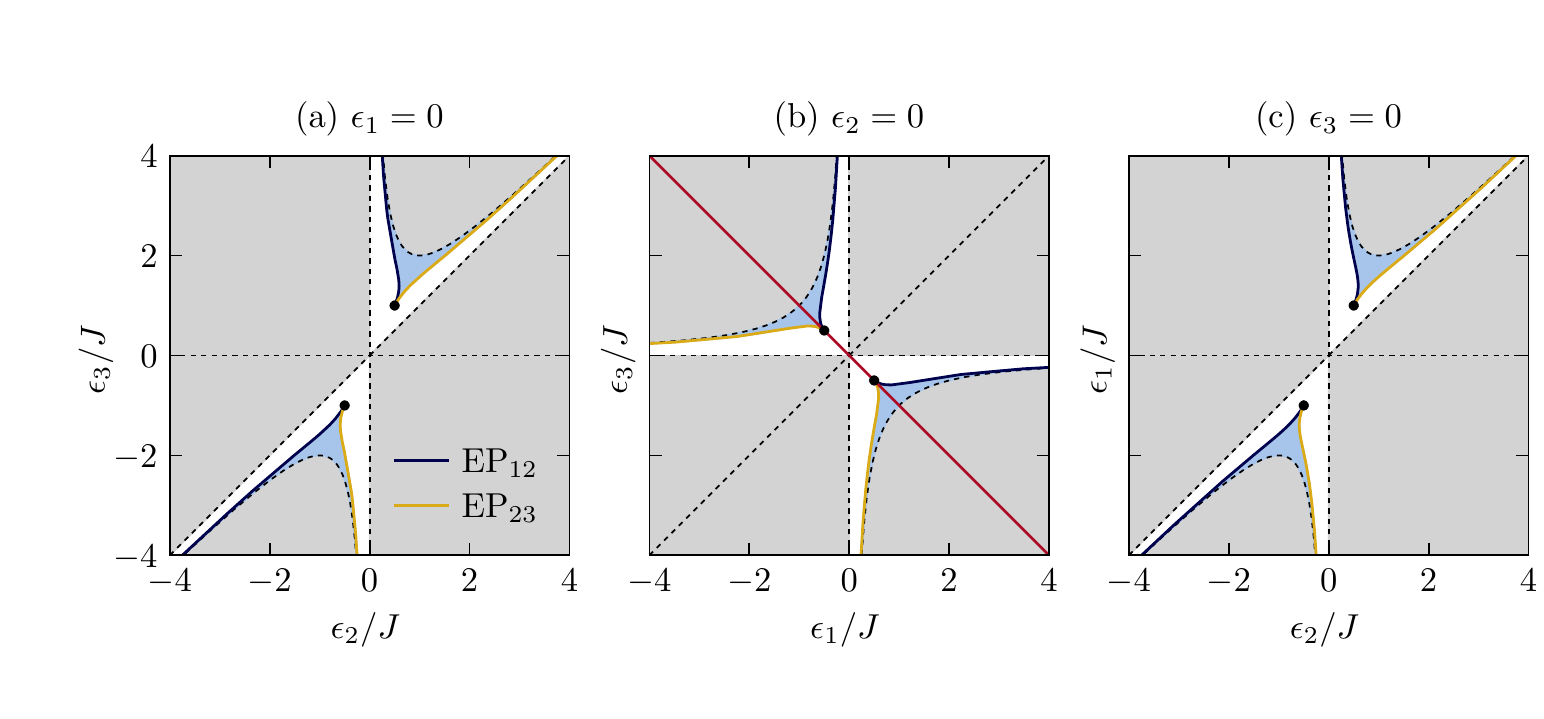}
  \caption{Slices along the coordinate planes in the $\qty(\epsilon_1, \epsilon_2, \epsilon_3)$ space. The grey shaded areas again indicate the parameter regions where no real solutions of \cref{eq:3_mode_gamma_0} exist. In contrast to \cref{fig:fig3} symmetric potentials ($\epsilon_1 = \epsilon_3$) and non-symmetric potentials are completely exclusive. Exceptional points occur between the ground and the first excited states ($\mathrm{EP}_{12}$) and between the first and second excited states ($\mathrm{EP}_{23}$), respectively. Their trajectories meet (marked by dots), which leads to a cusp catastrophe. Inside the cusps in the blue shaded area all three states are real while in the other areas only one real state is found. The solid straight line in (b) indicates the parametrisation used in \cref{fig:fig5}.}%
  \label{fig:fig4}%
\end{figure}

The regions in the $\qty(\epsilon_1, \epsilon_2, \epsilon_3)$ parameter space where solutions occur are shown in \cref{fig:fig4} for slices along the coordinate planes, which are symmetric in the sense that $\epsilon_1$ and $\epsilon_3$ are fully interchangeable. As in the two-dimensional case, an antisymmetric imaginary potential, i.\,e.\ $\gamma_1 = - \gamma_2$ and $\gamma_2 = 0$, is required by a symmetric real potential with $\epsilon_1 = \epsilon_3$ and arbitrary $\epsilon_2$. The spectrum for a symmetric real potential looks similar to the spectrum of the symmetric two-mode system in \csubref{fig:fig2}{a}, but with an additional state, which has zero energy everywhere. In contrast to the two-dimensional case, where the $\op{PT}$-symmetric potentials are special cases of symmetrised systems as shown in \cref{fig:fig3}, $\op{PT}$-symmetric and symmetrised systems are completely exclusive for non-trivial choices of the parameters as shown in \cref{fig:fig4}. That is, they only coincide for $\epsilon_1 = \epsilon_2 = \epsilon_3 = 0$.

Since all three states can now be potentially real, exceptional points occur in the spectrum where at least two states coalesce. The two independent second-order exceptional points between ground and first excited states and the first and second excited states are shown in \cref{fig:fig4}, respectively. They correspond to tangent bifurcations, where two stationary states coalesce and give birth to a pair of states with complex-conjugated energies. The trajectories of these exceptional points in the coordinate planes meet in the vicinity of the origin and create the characteristic form of a cusp. Hence, such a bifurcation scenario is called cusp catastrophe \cite{Poston1978}. The cusp points are again exceptional points but of third order, i.\,e.\ the coalescence of three states in a pitchfork bifurcation \cite{Demange2011,Gutoehrlein2013,AmShallem2015,Dizdarevic2015}. However, we will not discuss this phenomenon in detail here and instead refer the interested reader to \cite{Dizdarevic2015}, where we already reported on the occurrence of a cusp bifurcation in $\op{PT}$-symmetric Bose-Einstein condensates.

\begin{figure}
  \includegraphics[width=\columnwidth]{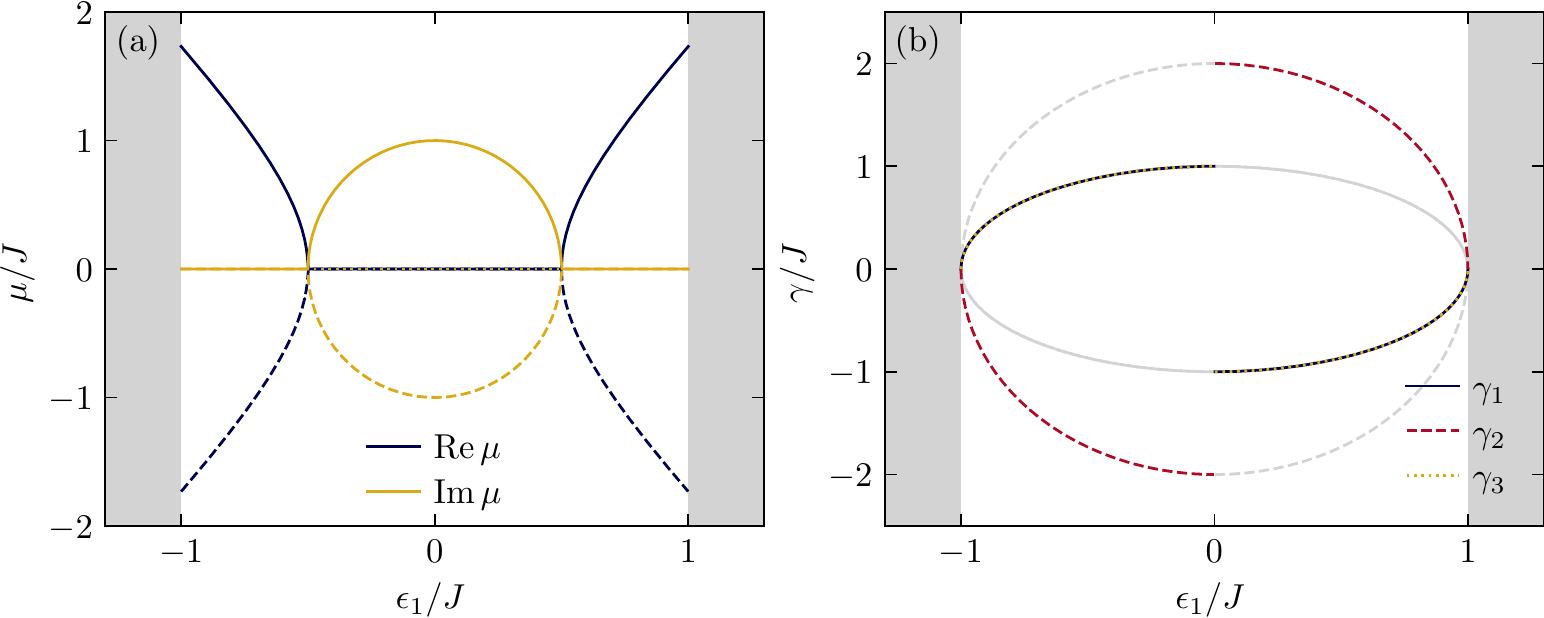}
  \caption{(a) Real and imaginary parts of the eigenvalues of the three-well system for an antisymmetric real potential with $\epsilon_2 = 0$ and $\epsilon_1 = - \epsilon_3$, which corresponds to the solid straight line in \cref{fig:fig4}. (b) Suitable gain and loss parameters are determined by \cref{eq:3_mode_solution_gamma_1,eq:3_mode_solution_gamma_2,eq:3_mode_solution_gamma_3} for $\gamma_0 > 0$. The same spectrum can be obtained with $\gamma_0 < 0$ being indicated by grey lines. The grey shaded areas indicate the regions where no symmetrisation is possible.}%
  \label{fig:fig5}%
\end{figure}

Note that there is a strong resemblance between \csubref{fig:fig4}{b} and the area in \cref{fig:fig3}, despite the fact that completely different parameter spaces are shown. This is, because the on-site energies and the gain and loss parameters differ only by an imaginary unit on entering the Schr\"odinger equation. Hence, real and imaginary parts of the potential are connected in a certain way. This is most obvious in $\op{PT}$-symmetric systems, where the real potential is symmetric and the imaginary potential is antisymmetric, corresponding to the first bisector in \csubref{fig:fig4}{b}. However, this relation also holds vice versa, that is if the imaginary potential is symmetric and the real potential is antisymmetric. Hence, we now choose an antisymmetric real potential with $\epsilon_1 = -\epsilon_3$ and $\epsilon_2 = 0$, which corresponds to the second bisector in \csubref{fig:fig4}{b}. Note that this type of potentials is partially embedded into the region of symmetrised Hamiltonians in \csubref{fig:fig4}{b} in the same way as the $\op{PT}$-symmetric potentials are in \cref{fig:fig3}. The spectrum along the second bisector is shown in \csubref{fig:fig5}{a} as a function of the on-site energy $\epsilon_1$, and the corresponding imaginary potential is chosen according to \cref{eq:3_mode_solution_gamma_1,eq:3_mode_solution_gamma_2,eq:3_mode_solution_gamma_3,eq:3_mode_gamma_0}, which is shown in \csubref{fig:fig5}{b}. Remarkably, the imaginary potential is always symmetric with $\gamma_1 = \gamma_3$. Such potentials are called anti-$\op{PT}$-symmetric \cite{Ge2013,Wu2014}, since the corresponding Hamiltonians satisfy
\begin{equation}
  \anticommutator{\op{PT}}{\op{H}} = 0
  \comma
\end{equation}
where the curly braces indicate the anti-commutator. $\op{PT}$ and anti-$\op{PT}$ symmetry are two special cases which show that symmetrised systems indeed bear certain symmetries, although these symmetries may not always be obvious.

Apart from $\op{PT}$ and anti-$\op{PT}$-symmetric configurations there again exist also completely asymmetric potentials, which still lead to symmetrised Hamiltonians. This holds, in particular, also for semi-symmetrised Hamiltonians. In principle one can find potentials with any number of real or complex-conjugated energies starting from one up to the number of wells. However, there is no straightforward way of constructing such semi-symmetrised Hamiltonians in multi-well potentials. Even in a three-mode system it is already feasible, though for three dimensions not yet being essential, to resort to numerical methods for calculating the potential parameters with a variational approach.

\section{Conclusions}%
\label{sec:conclusions}

In this paper we investigated the symmetrisation of Hamiltonians for few-mode systems. We introduced symmetrisation in a similar manner as quasi-Hermiticity. However, we focused on the bi-orthogonal nature of non-Hermitian quantum systems and their spectral properties. If a Hamiltonian is symmetrised, i.\,e.\ if \cref{eq:left_symmetrisation,eq:right_symmetrisation} hold, a pair of left and right symmetrisation operators exist. The corresponding spectrum then consists only of real or pairwise complex-conjugated eigenvalues. This is equivalent to quasi-Hermitian Hamiltonians if the kernels of the symmetrisation operators are empty, so that they are inverse to each other. We also showed, that a system can be semi-symmetrised if at least one eigenstate is in the kernel of the symmetrisation operators, so that the left and right symmetrisation operators are only semi-inverse to each other. The energies corresponding to such states are isolated spectral resonances.

We then applied the concept of symmetrisation to few-mode systems with complex potentials described by a discrete, linear Schr\"odinger equation. The symmetrisation of a two-mode system yields only $\op{PT}$-symmetric systems. However, the two-mode system is semi-symmetrisable, so that the ground state is always real. All few-mode systems with at least three dimensions are fully symmetrisable, so that there exist non-$\op{PT}$-symmetric Hamiltonians with all real spectra. A special class of such Hamiltonians shown here correspond to anti-$\op{PT}$ symmetry, where the imaginary potential is symmetric and the real potential is antisymmetric.

The concept of symmetrisation discussed in this paper can be applied, in principle, to most situations already well studied and understood in the presence of $\op{PT}$ symmetry. However, there is always a trade-off between the gain in generality and the loss in simplicity. On the one hand, the concept of $\op{PT}$ symmetry is simple and intuitive, which probably accounts for its success. On the other hand, as discussed above there are situations in which $\op{PT}$ symmetry is too restrictive and more general concepts are required. In particular, symmetrisation allows for more general and especially non-obvious symmetries. It allows for new applications as, for example, shown in \cite{Brandstoetter2019}, which are impossible with $\op{PT}$ symmetry.

Another example for an application is discussed in \cite{Lunt2017}, where the authors argue that asymmetric gain and loss is inevitable in real experiments. The authors also find further limitations by a stability analysis in a Bose-Einstein condensate with contact interaction described by a non-linear quantum system. Though the first system discussed has a symmetric real potential and is thus not symmetrised, Lunt \etal{} had to introduce an asymmetric real potential to stabilise their system. For zero contact interaction strength this corresponds exactly to our linear symmetrised two-mode system described in \cref{sub:two_mode_systems_with_arbitrary_gain_and_loss}. However, as mentioned in \cref{sub:semi_symmetrisation}, it is not possible to treat the non-linear system in a fashion similar to our treatment of the linear case. Hence, further numerical investigations are required to determine if and under which circumstances a single pair or multiple pairs of symmetrisation operators satisfying \cref{eq:left_symmetrisation,eq:right_symmetrisation} exist in the corresponding non-linear system.

Since the focus of this paper lies on the concept of symmetrisation itself, it would be interesting to investigate the phenomena occurring in symmetrised systems and their parameter dependencies in more detail in the future. This applies in particular to the exceptional points found in non-$\op{PT}$-symmetric parameter regions. In the spectrum shown in \csubref{fig:fig5}{a} two bifurcation scenarios emerge. They occur, however, not at the edge of the symmetrisable parameter region as in \cref{fig:fig3}, which would seem more intuitive. Instead the exceptional points are always in the middle of the regions shown in \csubref{fig:fig4}{b}.

It should further be noted that the occurrence of the borders in \cref{fig:fig3,fig:fig4} is clearly indicated by the parameters becoming complex numbers. Nevertheless, the spectra in \cref{fig:fig2,fig:fig5} show no sign of this whatsoever. It might thus be interesting to continue the parameters of the symmetrised matrix models analytically into the bi-complex plane to investigate the borders of the symmetrisable parameter regions.

Further, it seems crucial to extend the concept of symmetrisation to extended multi-well potentials, that is, going from a finite-dimensional to an infinite-dimensional Hilbert space. Since the matrix model discussed here is an approximation of a real, extended multi-well potential, which can, for example, be realised with ultra-cold Bose-Einstein condensates in a deep, one-dimensional optical lattice potential (see e.\,g.\ \cite{Kogel2019} and references therein), it is to be expected that at least a finite number of stationary states can be realised in asymmetric extended potentials. In \cite{Altinisik2019} we show examples of the application of symmetrisation to such extended double and triple-well potentials and found an excellent overall agreement between the finite and continuous models.

Another possible application for symmetrisation might be in many-body quantum systems described by quantum master equations \cite{Dast2014,Dast2016}. It was shown that the characteristic and dynamical properties of such systems correspond to the $\op{PT}$-symmetric models in mean-field approximation \cite{Dast2014}. However, the concept of $\op{PT}$ symmetry is not applicable directly to many-body quantum systems. Moreover, for a small number of particles a natural imbalance between gain and loss terms occurs. For these reasons it might be worthwhile to investigate many-body systems with respect to symmetrisation.


\ack
We thank M.\ Znojil and S.\ Rotter for fruitful discussions.




\section*{References}

\bibliographystyle{iopart-num}
\providecommand{\newblock}{}


\end{document}